# Monolithic Multi-parameter Terahertz Nano-micro Detector Based on Plasmon Polariton Atomic Cavity

Huanjun Chen[†], Ximiao Wang[†], Shaojing Liu, Zhaolong Cao, Jinyang Li, Hongjia Zhu, Shangdong Li, Ningsheng Xu\*, and Shaozhi Deng\*

State Key Laboratory of Optoelectronic Materials and Technologies, Guangdong Province Key Laboratory of Display Material and Technology, School of Electronics and Information Technology, Sun Yat-sen University, Guangzhou 510275, China

E-mail: stsdsz@mail.sysu.edu.cn; stsxns@mail.sysu.edu.cn.

[†]These authors contributed equally to the work.

**Abstract:** Terahertz signals hold significant potential for ultra-wideband communication and high-resolution radar, necessitating miniaturized detectors capable of multi-parameter detection of intensity, frequency, polarization, and phase. Conventional detectors cannot meet these requirements. Here, we propose plasmon polariton atomic cavities (PPAC) made from single-atom-thick graphene, demonstrating the monolithic multifunctional miniaturized detector. With a footprint one-tenth the incident wavelength, the detector offers benchmarking intensity-, frequency-, and polarization-sensitive detection, rapid response, and sub-diffraction spatial resolution, all operating at room temperature across 0.22 to 4.24 THz. We present the monolithic detection applications for free-space THz polarization-coded communication and stealth imaging of physical properties. These results showcase the PPAC's unique ability to achieve strong absorption and weak signal detection with a thickness of only $10^{-5}$ of the excitation wavelength, which is inaccessible with other approaches.

**Main Text:** Terahertz (THz: 0.1−10 THz) technology has been recognized as one of the "Top 10 technologies that will change the future of the world" (*1*). There is a continuing global effort to develop THz technologies to meet growing demands in areas such as ultra-wideband communication and high-resolution radar (*2−4*). Detectors are considered as one of the vital technologies for the THz communication and radar systems. In particular, with the advancement of information technology towards intelligence, the need for miniaturized detectors sensitive to multiple THz wave parameters, including intensity, polarization, frequency, and phase, has become increasingly pressing. Although such detectors have been widely explored in the visible to mid-infrared regions (*5−7*), a considerable performance gap persists between current technologies and demands in THz domain, especially beyond 1 THz. Commercially available coherent detectors rely on nonlinear electronic mixing effects, such as Schottky diodes, exhibiting high sensitivity and rapid speed at room temperature. They are capable of responding to amplitude, frequency, polarization, and phase of the THz wave. However, their operating frequency mainly is limited to below 1 THz, and the necessity of a local oscillator complicates their structure, hindering integration into large format arrays (*8*).

Photonic detectors, utilizing interband and intraband electron transitions in narrow bandgap semiconductors, offer benefits like sensitivity and fast response but demand cryogenic cooling for operation. Newly developed III-V high electron mobility transistor (HEMT) photonic detectors achieve high-sensitivity and swift detection at room temperature. However, their sensitivity degrades significantly beyond 1 THz (*9*, *10*), and to maintain high electron mobility for operation above 1 THz, cryogenic temperatures are necessary (*11*, *12*). Regarding thermal detectors, those functioning at room temperature demonstrate comparatively modest response times (ranging from microseconds to milliseconds) and moderate sensitivity (*13*, *14*). In contrast, high-sensitivity detectors like superconducting hot-electron bolometers necessitate cryogenic cooling (*15*, *16*). Significantly, owing to the weak THz wave absorption in contemporary photoactive materials, the coupling of THz signal to detectors is a challenge, necessitating antennas or cavities with tens to hundreds of micrometer footprints (*17*, *18*). Moreover, the detectors often demand combined optical components like polarizers, filters, and waveplates for multi-parameter detection.

In summary, the development of THz detectors currently encounters several challenges. On one hand, constrained by the optical absorption band gap of traditional semiconductor materials, the coupling efficiency between electromagnetic waves (EMWs) and carriers diminishes rapidly in the long-wavelength range. This reduction in coupling efficiency lowers the EMWs absorption efficiency and therefore photodetection sensitivity of the material. Particularly, when detector sizes are scaled down to the micro- and nanometer scale, overcoming the diffraction limit to achieve high EMW coupling with the micro-/nano-devices poses a significant challenge. On the other hand, the intrinsic absorption of EMWs by electrons within existing materials lacks selectivity in terms of frequency, polarization, and phase, making multi-parameter detection a formidable task. Typically, the current approach involves employing a heterogeneous integration method of "detector + optical components" to combine detectors with polarizers, filters, and phase modulators. However, this integration method introduces issues such as high system complexity, large size, and limited bandwidth, thereby constraining devices of miniaturization, high integration, high speed, and lightweight. Apparently, to advance the development of THz detectors, it is crucial to explore principles beyond the traditional photoelectric effects. This involves concentrating free-space THz waves into micro-/nanomaterials and facilitating efficient conversion of THz wave energy into electrical signals with multi-parameter sensitivity.

Van der Waals (vdW) two-dimensional (2D) crystals with atomic thicknesses, represented by monolayer graphene, can sustain diverse types of charged elementary excitations, enabling them to host a plethora of polaritonic modes (*19*) with high field confinement and low loss, especially in the THz and mid-infrared spectral regions. In particular, monolayer graphene with hexagonally arranged carbon atoms in a 2D plane exhibits gapless-linear band dispersion, rendering the existence of 2D Dirac plasmon polariton (PP) spanning the ultraviolet to THz and even microwave regions (*20*). However, this unique property has not been explored for detection. Our current study shows that structuring monolayer graphene into regular structures can give rise to plasmon polariton atomic cavities (PPAC) exhibiting PP resonances with strong electromagnetic field confinement in the THz domains. The atomic cavity enables strong absorption of THz waves that is necessary to detection. The resonance frequency of the PPAC can be effectively tuned across a broad spectral range by either modifying the geometry of individual PPAC or arranging a variety of PPACs into ordered arrays. This means that one may decide on a prioritized frequency for detection through a design of the cavity. These unique characteristics provide opportunities for exploring materials and devices with functionalities otherwise absent in their conventional counterparts.

In this study, we present an approach for creating ambient-condition THz detectors using an interconnected array of the PPACs, which convert the THz EMWs into electrical signals from distinct morphology-dependent PP resonances spanning from 0.5 THz to 10 THz. The resonances and thus the signals are highly sensitive to THz polarization in geometrically anisotropic PPACs. In the photoactive channel, the interconnected PPAC array functions as both an electromagnetic absorber and a carrier transport layer. When exposed to THz waves, the PP resonances are excited and subsequently decay into hot carriers, which are efficiently transported to and collected by the electrode. Moreover, a single atomic layer thickness of the PPACs ensures low dark current in the detector. These unique properties enable room-temperature sensitive and rapid THz direct detection with benchmarking frequency and polarization-resolved attributes. Importantly, the PPACs eliminate the need for an antenna or an additional light-absorbing layer, resulting in a significantly smaller footprint compared to the incident wavelength. This capability enables THz detection below the diffraction limit, opening the door to the development of highly integrated, high-resolution intelligent focal plane arrays in the THz domain, where such devices are currently rare. We additionally demonstrate, for the first time, the first monolithic detection applications for free-space THz polarization-coded communication and stealth imaging of physical properties, achieving spatial resolution approaching the diffraction limit. These advancements are all made possible by the principle of PPACs and the corresponding detectors.

It is noted that the PPAC principle can be extended to various polaritonic materials, enabling the design and development of miniaturized photonic and optoelectronic devices across broad spectral ranges using the PPAC framework.

**Principle and design of the PPAC**

Central to the PPAC is micro- or nanostructures characterized by regular shapes and a thickness limited to a single atomic layer (Fig. 1, A and B). The interaction between the electromagnetic field and charged elementary excitations in the atomic layer gives rise to polaritons. Our theoretical calculations demonstrate that these polaritons within the PPAC can establish robust localized resonances, resulting in pronounced absorption strongly dependent on incident frequency and polarization. Furthermore, organizing individual PPACs into ordered arrays (Fig. 1A, right panel) allows for the fine-tuning of their electromagnetic responses, encompassing absorption intensity and resonance frequency. This tuning is achieved by harnessing the near-field coupling between localized polariton resonances within each PPAC.

Taking graphene, a monolayer carbon atom network, as an example. The monolayer graphene, with a gapless band structure and sufficient high doping, can sustain PPs propagating as isotropically evanescent waves within the 2D plane. The PP dispersion, *i.e.*, $\omega(q)$, where $q$ is the wavevector, in homogeneous graphene is highly nonlinear, with $q$ much larger than that of free-space EM field at the same frequency (fig. S1A). This places a significant challenge for exciting the PP directly from the far-field, and consequently greatly hinder the practical applications. Such a large wavevector mismatch can be circumvented by introducing the PPACs. In a PPAC, PPs couple with the incident THz wave, propagate within the atomic layer, encounter the atomically sharp boundaries, and are then reflected by these boundaries. In this scenario, the incident THz wave acts as the driving force, while the atomic boundaries and the self-consistent field created by all the electrons provide the restoring forces, resulting in localized PP resonances in the PPAC. These resonance modes can be described through a non-Hermitian Hamiltonian that accounts for both the radiative and non-radiative decays of the PPs,

$$H\Psi_m = \begin{pmatrix} 0 & -\nabla\times & \hat{\beta}\delta(z) \\ -\nabla\times & 0 & 0 \\ \hat{\beta}\delta(z) & 0 & -\tau^{-1}\delta(z) \end{pmatrix} \begin{pmatrix} \tilde{E}_m \\ \tilde{H}_m \\ \tilde{U}_m \end{pmatrix} = j\tilde{\omega}_m \begin{pmatrix} \varepsilon & 0 & 0 \\ 0 & -\mu_0 & 0 \\ 0 & 0 & -\delta(z) \end{pmatrix} \begin{pmatrix} \tilde{E}_m \\ \tilde{H}_m \\ \tilde{U}_m \end{pmatrix} \quad (1)$$

where $\tilde{\omega}_m$, $\tilde{E}_m$, $\tilde{H}_m$, and $\hat{\beta}\tilde{U}_m$ are the complex eigen-frequency, electric field, magnetic field, and surface current within the PPAC, respectively. The function $\hat{\beta}$ describes the cavity geometry of the PPAC. Parameters $z$, $\tau$, $\varepsilon$, and $\mu_0$ stand for the coordinate along z-axis, electron relaxation time of graphene, permittivity of graphene and permeability of vacuum. The THz absorption/reflection spectra and the localized electromagnetic field distribution of a PPAC with specific shape and size can readily be obtained by solving Eq. 1 (supplementary text ST1).

We compare the THz absorption spectra of unpatterned graphene (dashed line, Fig. 1B) with that of a circular graphene nanodisk with a diameter ($D$) of 10 μm (blue solid line, Fig. 1B). The graphene layers have a Fermi energy ($E_F$) of 0.3 eV, a common value for graphene supported on dielectric $SiO_2$ subsitrates (*21*, *22*). Under linear polarized excitation, the graphene disk exhibits a prominent resonance at 1.24 THz, resulting in significantly enhanced and localized EM fields within the disk (Fig. 1C, inset), in contrast to the unpatterned graphene (fig. S2). Such resonance is of dipolar nature, as evidenced by the charge distribution patterns (fig. S3, movies S1 and S2). The $\omega_{res}$ can be described as $\omega_{res} = \sqrt{-\frac{e^2 E_F}{4\pi\varepsilon_0 \bar{\varepsilon} \hbar^2 \eta_1 D}}$ under the assumption that $D$ is much smaller than the incident wavelength (Supplementary text ST1). Therefore, the $\omega_{res}$ can be adjusted by altering either the diameters of the graphene disks or the $E_F$, as evidenced by the THz absorption spectra (Fig. 1B, colored lines, fig. S4). Notably, $\omega_{res}$ exhibits a nonlinear decease against the disk diameter (Fig. 1C), a behavior captured by a square-root scaling law expressed as $\omega_{res} \propto D^{-1/2}$. The distinct square-root scaling of $D^{-1/2}$ stands in stark contrast to that observed in conventional noble metal plasmonic nanoparticles and Mie nanoresonators constructed from high-refractive-index nanoparticles, where the scaling typically involves either $D$ or $1/D$ (*23*). This results in a spectral tuning range spanning from under 0.5 THz to over 5 THz as $D$ changes from 60 μm to 1.3 μm (Fig. 1C). Importantly, unlike traditional semiconductors whose EMW absorption is governed by band-to-band electron transition with bandwidths restricted by the electron bandgap (Fig. 1D), the PPAC exhibits tunable EMW absorption covering the broad THz band from 0.5 THz to 10 THz (Fig. 1B and Fig. 1D), owing to the gapless electronic structure of monolayer graphene and the excitation of the adjustable PP resonances. Furthermore, by delicately arranging the PPAC into an array with a duty ratio of 0.5, the absorption maximum can be as high as 22.5%, with only a monolayer thickness of 0.7 nm and regardless of the resonance frequency (Fig. 1B). This value is much higher than those of traditional semiconductor films with the same thickness (Fig. 1D), highlighting the significant potential of PPAC arrays in developing miniaturized, high-performance THz photonic and optoelectronic devices where EMWs absorption plays a crucial role.

The $\omega_{res}$ can also be tailored by utilizing PPACs of varied shapes but with comparable feature sizes (Fig. 1E). More important, the absorption spectra demonstrate remarkable sensitivity to the polarization of the THz wave when an anisotropic nanoribbon is inspected. The highest and lowest absorption occur when the polarization of the THz wave aligns along the longitudinal and transverse direction of the nanoribbon, respectively (Fig. 1F). The resonance's absorption intensity varies in accordance with a cosine function as the incident THz polarization changes (Fig. 1F, inset).

It is worth noting that near-field plasmonic coupling can occur between neighboring PPACs, leading to an additional enhancement of the electromagnetic field within the gaps between them. This coupling provides a means to tune the $\omega_{\text{res}}$ by arranging various PPACs into ordered arrays, as expressed by $\omega_{\text{res}} = \sqrt{-\dfrac{e^2 E_F}{4\pi\varepsilon_0 \bar{\varepsilon}\hbar^2 \eta_1 D} - \dfrac{c\bar{n}g}{2\pi\bar{\varepsilon}\sqrt{S}}\Gamma_{\text{rad}}}$ (Supplementary text ST1) (*24*, *25*), with parameters *g* and *S* determined by the symmetry of the array and gaps between adjacent PPACs, respectively. Clearly, the resonance exhibits a redshift behavior as the gap size is reduced (fig. S1, B and C). In principle, the theoretical model we developed for PPACs is inherently extendable to diverse atomic polaritonic materials. Examples include transition metal dichalcogenides and α-MoO$_3$, which support exciton polaritons in visible spectral range and phonon polaritons in mid-infrared to THz spectral range, respectively. This facilitates the design and development of photonic and optoelectronic devices within the PPAC framework across a wide range of spectral regions. The choice of polaritonic materials with suitable dielectric functions, aligned with the desired frequency range, enables versatility in device engineering.

**Fabrication and characterization of the PPACs**

The PPACs were fabricated by etching a monolayer graphene flake, which was transferred onto a silicon substrate covered with a 300 nm-thick $SiO_2$ layer (MM1). Our current technology has reached the level of wafer-level processing of 4-inch PPACs (Fig. 2A). Scanning electron microscope (SEM) images of representative PPACs with different shapes and sizes are shown in Fig. 2, A to E. The fabrication process did not compromise the atomic thickness of the PPACs, as confirmed by atomic force microscope and high-resolution transmission electron microscope characterizations (Fig. 2, F and G). The $\omega_{res}$ can be tailored by adjusting the diameters of the graphene disks, as evidenced by both theoretical (Fig. 2H, solid lines) and experimental (Fig. 2H, symbols) THz absorption spectra. The $\omega_{res}$ extracted from the absorption spectra for different disk diameters exhibits a close alignment with the dependency pattern on disk diameter as anticipated by theoretical calculations (Fig. 1C). The absorption spectra allow for evaluating the Q-factor of the PP resonance, which is defined as $Q = \omega_{res}/\Gamma$, with $\omega_{res}$ and $\Gamma$ the frequency and linewidth of a specific resonance (supplementary text ST2). The $\omega_{res}$ is a function of disk diameter (Fig. 1C). The measured Q-factors increase monotonically against $\omega_{res}$, which is consistent with the theoretical results (Fig. 2I). This phenomenon can be attributed to the fact that for a graphene cavity array with a specified duty ratio, the polariton dissipation rate, encompassing both radiation ($\Gamma_{rad}$) and absorption ($\Gamma_{abs}$) rates, remains constant (24). The $\Gamma$ is the sum of $\Gamma_{rad}$ and $\Gamma_{abs}$. As a result, the Q-factor of the graphene cavity array scales linearly with the resonant frequency. The Q-factor of the PPAC is moderate within the investigated frequency range, due to the relatively broad resonance caused by the high dissipation rate of the PP resonances. However, it is important to note that such rapid dissipation significantly accelerates the decay of PPs into hot electrons, facilitating subsequent THz detection, as discussed below.

PPACs of ribbon shapes enable polarization-sensitive THz absorption, with the absorption maxima adjusting in accordance with a cosine function as the incident THz polarization changes (Fig. 1F and Fig. 3E). The ability to tune the resonance frequency and achieve polarization-sensitive absorption renders PPACs and their ordered arrays highly suitable for frequency- and polarization-resolved THz detection. These are completed monolithically.

**Room-temperature THz detector based on PPAC**

Central to the detector design are two critical elements: firstly, addressing the discrepancy between the broad spot size of the incident THz wave and the confined device area. This involves effectively concentrating the THz wave with specific central frequency and polarization onto the photoactive layer, thereby enhancing electromagnetic absorption; secondly, the efficient conversion of absorbed electromagnetic energy into current or voltage signals, subsequently gatherable by external circuits. To this end, we introduce a detector configuration comprising a photoactive channel connected by two metal electrodes (Fig. 3A). The channel layer consists of a monolayer graphene, with one half composed of interconnected PPACs through graphene strips. The PPACs can localize and absorb the THz wave (Fig. 3A) and then generate hot electrons, while the strips serve as pathways for carrier transport. The remaining half is left unpatterned to facilitate charge transportation as well. The THz detectors are supported onto a pristine silicon wafer covered with a 300-nm-thick $SiO_2$ layer (MM2). The presence of the $SiO_2$ layer induces p-type doping of the graphene, resulting in an $E_F$ of ~ 0.3 eV (Supplementary text ST3) (*26*). The SEM image in Fig. 3A (lower panel, left) depicts a representative detector comprising disk-shaped PPACs, with 25 PPACs arranged in a 5 × 5 array, spaced 2 μm apart. The PPAC diameter and the graphene strip width are 30 μm and 5 μm, respectively. The overall dimensions of the detector are 155 × 335 μm², with a resistance of 1.5 kΩ (MM3, fig. S5).

Leveraging their PP resonances, the PPACs efficiently concentrate incoming THz waves into the photoactive channel layer and concurrently act as absorbing elements. Consequently, the need for additional antenna structures or EMW absorption layers is obviated, streamlining the detector's design and enhancing its compactness. Upon excitation, the resonances undergo intraband Landau damping due to their small resonance energy. Additionally, the PPs will be scattered by the edges of the graphene nanodisk, leading to the generation of hot carriers with an elevated temperature. This mechanism facilitates efficient room-temperature THz wave detection. In our proposed device, when exposed to THz waves, the PPAC region experiences a frequency-dependent temperature increase $\Delta T$ (fig. S6, black line), corresponding well to the THz absorption spectrum of the PPAC (fig. S6, red line). Consequently, temperature and hot-carrier-density gradients arise across the PPAC and unpatterned graphene areas (Supplementary Text ST4, fig. S7).

The interplay of these factors collectively contributes to the hot-carrier photo-thermoelectric effect (HCPTE), inducing a directional flow of carriers along the channel. This allows for the electrical readout of hot carriers under zero bias, as illustrated in Fig. 3A. Moreover, the fabrication process introduces a minor number of defects into the PPACs (fig. S8), which facilitate the decay of PP resonances into hot carriers, thereby enhancing subsequent THz detection (*27*, *28*). The resulting photo-induced voltage *V* can be expressed as,

$$V = \int_0^L S(x)\frac{dT(x)}{dx}dx \tag{2}$$

where *L* is the channel length, *S*(*x*) is the Seebeck coefficient within the channel, and *T*(*x*) is the local temperature. The bias-free HCPTE mechanism, in conjunction with graphene's high carrier mobility and the short conducting channel, offers the potential for efficient and rapid THz detection with small dark current at room temperature.

The detector's performance was assessed at room temperature through photo-induced voltage measurements under THz illuminations spanning frequencies from 0.22 THz to 4.24 THz, output from a gas laser (MM3, fig. S9A). A comparison of the current−voltage (*I*−*V*) curves of the detector with and without THz illumination reveals a distinct rise in the *I−V* curve upon THz excitation, maintaining a consistent slope (fig. S5). Remarkably, measurable current is observed even at zero bias voltage, confirming the dominance of the HCPTE effect in the THz photoresponse, enabling zero-bias operation of the PPAC detector. Additionally, when subjected to 0.34-THz illumination at 563 µW, an induced temperature variation of $\Delta T = 30$ mK generates a photo-induced voltage of 2.1 µV according to Eq. (2) (Supplementary text ST4), agreeing well with the experimental measurement (2.0 µV). The photo-induced voltage demonstrates a linear dependence on incident THz power (Fig. 3B), resulting in a responsivity of $R_v = 2.48$ V/W, referenced to the power absorbed in the PPAC region. It is noted that referencing responsivity to absorbed power can sometimes result in an overestimation of detector sensitivity. This stems from the difficulty faced by conventional devices in focusing THz light into a photoactive area much smaller than the incident wavelength (*29*). However, our proposed detector benefits from the efficient confinement of incident THz waves by PPAC resonances, rendering it reasonable to assess our device's responsivity with respect to the absorbed power of the PPACs.

To evaluate the spectral response of the detector, we measured its responsivity at zero bias for various THz frequencies ranging from 0.22 THz to 4.24 THz (fig. S9B). As depicted in Fig. 3C (orange), the PPAC detector demonstrates a pronounced frequency-dependent photo-induced voltage generation, peaking at 0.70 THz. This trend is consistent with the calculated absorption spectrum of the PPAC with a 30 μm disk diameter, which sustains a PP resonance at 0.73 THz. Minor deviations can be attributed to experimental sample variations in disk diameter. Notably, the detector's operating frequency is tunable by adjusting the disk diameter (Fig. 3C); PPACs with diameters of 4 μm, 7 μm, and 78 μm exhibit peak responsivity at 2.52 THz, 1.84 THz, and 0.50 THz, respectively. These findings clearly highlight the frequency-sensitive characteristic of our proposed THz detector. At the resonance frequencies, the on/off ratio of the detector at a power density of 100 mW/cm$^2$ reaches 100 ~ 150, meeting the performance requirements of radar imaging. Meanwhile, the detector exhibits a responsivity bandwidth, defined as the full-width at half maximum (FWHM) of the photo-induced voltage, of approximately 1 THz.

The detector exhibits inherent polarization-resolved capability by leveraging the anisotropic absorption of the ribbon-shaped PPACs. For a detector consisting of a 6 × 5 array of nanoribbons (width × length = 15.0 μm × 5.0 μm), the photo-induced voltage upon 0.34-THz illumination exhibits a pronounced dependency on the incident THz polarization (Fig. 3E, blue symbols). Specifically, the photo-induced voltage reaches its peak when the THz polarization aligns with the longitudinal direction of the ribbon and decreases to a minimum when the polarization is rotated by 90°, corresponding to the transverse direction of the nanoribbon. Furthermore, the variation of the photo-induced voltage in response to THz polarization aligns with the simulated (Fig. 3E, red dashed line) and measured (Fig. 3E, green symbols) changes in the absorption intensity of the graphene nanoribbon under different incident polarizations. Such agreement verifies that the polarization-sensitivity of the detector is originated from the anisotropic absorption of the PPACs rather than from the metallic electrodes (fig. S10). The variation of the photo-induced voltage in response to the polarization angle can be accurately fitted using a cosine square function, allowing for calculating the polarization ratio (PR). This ratio, which is defined as the division between the maximum and minimum photo-induced voltage values for different illumination polarizations, is calculated to be 63. This value surpasses most THz detectors that rely on 2D crystals integrated with antennas and metamaterials (Supplementary Table S1).

In addition to their polarization sensitivity, the resonance frequency of ribbon-shaped PPACs can be conveniently adjusted by modifying their aspect ratio (AR), which is defined by dividing the length by the width of the ribbon. This degree of freedom empowers us to design compact detectors using ribbon-shaped PPACs that are customized for a specific illumination frequency, thus enabling the miniaturization of devices (Supplementary ST1, fig. S11). For instance, if we choose an AR of 5, a graphene nanoribbon with dimensions of 2.5 μm in length and 0.5 μm in width exhibits a PP resonance at 2.52 THz (fig. S11). By arranging 30 of these nanoribbons into a 6 × 5 array, a detector can be created with a photoactive area measuring $11.0 \times 19.0$ μm$^2$ (Fig. 3A, lower panel, right). This detector shows remarkable polarization-selective photo-induced voltage responses (fig. S12), boasting a responsivity of 12.7 V/W and a PR of 88 at 2.52 THz. The footprint of the detector is only 1/10 of the incident terahertz wavelength (119 μm), indicating that it is capable of operating at the deep-subwavelength scale. Such a small device can effectively suppress room-temperature noise amplitude (fig. S13), resulting in a noise-equivalent power (NEP) of 2.8 nW/Hz$^{0.5}$.

The HCPTE mechanism, combined with the detector's small footprint and high carrier mobility of graphene, enables high-speed detection performance. The temporal response of the detector corresponding to Fig. 3D was characterized by illuminating it with an ultrafast THz pulse (0.1 to 3 THz) and recording the temporal electrical response of the device (see MM3, fig. S14). The photo-induced voltage follows the pulse THz excitation and has a rise/fall time of 4.5 ns/7.2 ns (Fig. 3D). It is important to note that in principle, the characteristic response time of a graphene detector based on the HCPTE mechanism can reach the order of tens of picoseconds (*30*). This indicates that there is ample of room for speeding up our detector, which can be achieved through various approaches such as substituting high-resistance substrates, reducing channel lengths, introducing asymmetric electrodes, and enhancing the carrier mobility of graphene (*31*, *32*).

The small photoactive area of the PPAC detector can extend into the deep subwavelength region in the THz domain, enabling detection with high spatial resolution. For instance, considering the focal length (101.6 mm) of the lens utilized in our study, a 2.52-THz (λ = 119 μm) laser beam, with a diameter of ~35 mm, can be focused to a diffraction-limited spot with a FWHM of ~390 μm (Supplementary text ST5). By raster scanning the detector across the focal point (MM4 and fig. S15), the focused spot can be spatially mapped out. The results reveal a distinct spot pattern (Fig. 3F), where intensity differences between adjacent positions, separated by 25 μm, are discernible (Fig. 3G). To the best of our knowledge, this value, which is only ~1/5 of the incident wavelength, represents the most significant achievement to date in realizing THz wave detection capabilities beyond the diffraction limit, utilizing frequencies of 2.52 THz and below. This highlights the exceptional capability of our detector technique for achieving detection with spatial resolution that surpasses the diffraction limit. It is anticipated that the spatial resolution of detection may be 6 μm if a 10-THz source is employed. These attributes position our detector for focal-plane imaging of practical objects with high speed and high spatial resolution.

**Application demonstrations of the PPAC THz detectors**

With its compact size, sensitivity to polarization and frequency, rapid response, and high spatial resolution, the PPAC detector opens up opportunities for THz domain applications that are currently beyond the capabilities of conventional commercial detectors. Firstly, the miniaturized PPAC detector with a high on/off ratio can facilitate THz stealth imaging with a high spatial resolution. To that end, THz image of a metallic plate with a hollow bowtie structure at its center, which is placed inside an envelope, is captured by illuminating the plate with an expanded 2.52-THz beam and raster-scanning the detector across the image plane (Fig. 4, A and B, MM4). The gap between the two tips of the metallic plate measures 68.9 μm (Fig. 4B, inset, SEM image), which is comparable to half of the THz wavelength ($\lambda/2 = 59.5$ μm). The hollow bowtie region is clearly discernible in the image (Fig. 4C). Notably, based on the FWHM of the photo-induced voltage line profile near the gap region, it is evident that regions with a separation of 475 μm can still be resolved (Fig. 4C and the inset line profile). This indicates that the spatial resolution of the detector for application in THz stealth imaging of practical objects can be as high as ~500 μm. Such spatial resolution of THz stealth imaging has not been found in literature. In contrast, the hollow metallic plate cannot be resolved (Fig. 4D and the line profile shown inset) when employing commercially available room-temperature THz detectors with a photoactive area of 3 × 3 mm² (fig. S16). It is noteworthy that in our current study, high-resolution imaging was achieved solely through the compact size of the detector, without optimizing or designing the imaging optical system. The other components in our imaging system give rise to problems to break the diffraction limit. Subsequent investigations may extend imaging capabilities by incorporating vector beam illumination and integrating algorithmic reconstruction alongside other advanced technologies in an imaging systems.

The polarization-selectivity of the PPAC detector, echoing with its small footprint, holds promise for monitoring the physical properties, *e.g.*, the strain distribution, of concealed objects. To analyze strain distribution is important, for example, in a silicon wafer, integrated circuit chips, and nanodevices. Specifically, strains in a thin film can induce changes in the refractive index along the direction of the applied strain. Consequently, this alteration can affect the polarization state of THz waves as they pass through the film. These modified waves can then be detected by our polarization-sensitive detectors, facilitating the perception and identification of the induced stress (*33*, *34*). To visually convey this application scenario, a 2.52-THz laser with a linear polarization angle of 45° was directed onto a polyester film. This film was enclosed within an envelope and raster-scanned within the focal point of the beam. Mechanical strain was then applied to induce deformation of the polyester film (Fig. 4E). In contrast to the undistorted film (Fig. 4F, left panel), the compressed polyester film exhibited a non-uniform distribution of higher photo-induced voltage (Fig. 4F, right panel). To further confirm that the non-uniform photo-induced voltage distribution is associated with the strain applied to the polyester film, numerical simulations were conducted to analyze the strain distribution in the polyester film (Supplementary text ST6). Upon applying a compression force perpendicular to the left and right edges of the film, the simulated strain of the film exhibits a significantly non-uniform distribution compared to the case without stress (fig. S17). This simulated strain distribution aligns with the experimental photo-induced voltage distribution in Fig. 4F. Interestingly, according to the simulation result, the strain in the middle region of the film is relatively small, attributed to the mutual cancellation of the tensile force on the upper surface and the compressive force on the lower surface. This feature corroborates well with the photocurrent mapping shown in Fig. 4F (right panel), suggesting that different strains will generate opposite changes of the refractive index. This outcome distinctly demonstrates, for the first time, the ability to detect the strain of concealed materials in THz stealth imaging. Furthermore, it reveals that variations and distributions of physical properties resulting from changes in refractive index can now be detected in THz stealth imaging systems using our detector.

The PPAC can also enable the polarization-coded THz wireless communication, where information is encoded using two orthogonal polarizations representing "1" and "0" states, respectively. As a proof-of-concept, the sequence "SYSU" was initially converted into American Standard Code for Information Interchange (ASCII) codes. Subsequently, it was modulated onto the dual polarization states of a 2.52-THz carrier (Fig. 4G, MM5), with a polarization angle of 0° for "1" and 90° for "0". The encoded data is then transmitted through free space by the carrier and captured by the PPAC detector. The photo-induced voltage responses of the detector correlate precisely with the original encoded information (Fig. 4H), showcasing the potential of our device as a receiver for future THz wireless communication applications. This result is also the first demonstration with a carrier signal up to 2.52 THz; a very large wide-band frequency. Without the integration of polarizing optics in front of the PPAC detector, the receiver system can be considerably more compact than commercial THz detectors lacking inherent polarization selectivity. This indicates that our detector may be a favorable choice for used in a receiver chip in future THz telecommunication.

**Concluding remarks**

Developing a compact nano-micro multifunctional THz detector remains a significant challenge. Traditional THz detection materials, which rely on interband/intraband transitions for optical absorption, have constrained and inefficient absorption in the THz spectral range. This limitation hinders their ability to selectively respond to the frequency, polarization, and phase of electromagnetic waves. Therefore, it is crucial to explore new principles for creating miniaturized, multifunctional THz detectors.

In the current study, we introduce the principle of PPAC, which we believe is absolutely necessary for addressing the above challenge. The PPAC, formed by structuring a gapless atomic graphene monolayer into regular micro- or nanostructures, induces PP resonances, enabling strong EMWs absorption across a broad spectral range, which conventional semiconductors cannot achieve. These resonances concentrate THz waves into a much smaller photoactive area, eliminating the need for cumbersome antennas or metamaterial layers. The photoactive channel allows for efficient electrical readout, facilitating direct plasmon polariton-to-electron conversion through the HCPTE effect, resulting in rapid THz detection at room temperature. Furthermore, the geometry-dependent plasmon polariton resonances enable the benchmarking ability of the detector to be both polarization- and frequency-sensitive across a wide spectral range spanning from 0.22 to 4.24 THz.

Secondly, our detector is the first to offer a monolithic solution for a multifunctional nano-micro device, eliminating the need for auxiliary components such as antennas and lenses. To the best of our knowledge, this represents the most efficient approach for realizing a compact, multifunctional THz detector. This innovation allows for significant miniaturization, with the detector's footprint being merely one-tenth of the incident THz wavelength. Our device technique provides a practical solution for monolithic chip fabrication.

Thirdly, these miniaturized detectors, distinguished by their multi-parameter and rapid detection capabilities across a broad THz spectral range, position themselves as superior in applications surpassing current state-of-the-art THz detectors (Supplementary Table S1). Specifically, achieving a detection spatial resolution of 25 μm, which is only 1/5 of the incident wavelength, exceeds the diffraction limit and stands as the highest value reported in THz detectors to date. This capability enables the detector to effectively perform stealth imaging of physical properties with high spatial resolution. For example, it can image strain distributions in a plastic plate, as demonstrated here. As a result, one can now not only see through black clothes but also discern their properties. Additionally, the pioneering utilization of a 2D crystal-based detector for chips in THz polarization-coded wireless communication is demonstrated, leveraging the polarization-resolved capability of the PPAC principle and detector. Utilizing polarization-coded technology offers a promising avenue for enhancing the transmission rate and capacity of THz wireless communications. This becomes particularly significant when polarization multiplexing technology incorporates grayscale resolution, rather than merely discerning the two polarization states, as it enables even greater improvements in transmission rate and capacity.

In principle, the PPAC concept is extendable to other 2D crystals with polariton properties, such as α-MoO$_3$, black phosphorus, and α-V$_2$O$_5$ (*35−38*). This implies the possibility of developing a variety of detectors operating across a broad spectral range, from THz to visible wavelengths, by selecting materials with suitable polariton bands within that spectrum. In summary, the combination of high-performance multi-parameter detection and the compact size of the PPAC positions it as a new paradigm for the design and advancement of cutting-edge terahertz detectors and chips, holding the potential to redefine future imaging and wireless communication techniques.

**References and Notes**

1. G.T. Huang, *Technology Review* **107**, 32 (2004).
2. B. Ferguson *et al., Nat. Mater.* **1**, 26−33 (2002).
3. S. Koenig *et al., Nat. Photonics* **7**, 977−981 (2013).
4. H. Matsumoto *et al., Nat. Electronics* **3**, 122−129 (2020).
5. Y. Fan *et al., Nature,* 1−7 (2024).
6. Q. Guo *et al., Nat. Mater.* **17**, 986−992 (2018).
7. J. Wei *et al., Nat. Photonics,* **17**, 171−178 (2023).

8. K. Bourzac, *Nature* **483**, 388 (2012).

9. M. Bauer *et al., IEEE Trans. Terahertz Sci. Technol.* **9**, 430−444 (2019).

10. A. Rogalski *et al., J. Infrared, Millimeter, Terahertz Waves* **43**. 709−727 (2022).

11. H. Qin *et al., Appl. Phys. Lett.* **110**, 17 (2017).

12. W. Knap *et al., J. Infrared, Millimeter, Terahertz Waves* **30**, 1319−1337 (2009).

13. S. Cherednichenko *et al., IEEE Trans. Microwave Theory Tech.* **1**, 395−402 (2011).

14. R. Müller *et al., J. Infrared, Millimeter, Terahertz Waves* **36**, 654−661 (2015).

15. B. S. Karasik *et al., IEEE Trans. Terahertz Sci. Technol.* **1**, 97−111 (2011).

16. S. Seliverstov *et al., IEEE Trans. Appl. Supercond.* **25**, 1−4 (2014).

17. S. Chai *et al., IEEE Microw. Wireless Compon. Lett.* **24**, 869−871 (2014).

18. J. Zmuidzinas *et al., Proceedings of the IEEE* **92**, 1597−1616 (2004).

19. D. N. Basov *et al., Science.* **354**, aag1992 (2016).

20. A.N. Grigorenko *et al., Nat. Photonics* **6**, 749−758 (2012).

21. N. K. Emani *et al., Laser Photonics Rev.* **9**, 650−655 (2015).

22. D. Yoon *et al.*, *Phys. Rev. B* **80**, 125422 (2009).

23. C. F. Bohren *et al.*, *Absorption and scattering of light by small particles, Wiley-Interscience, New York*. (1983).

24. H. Zhu *et al., Front. Mater.* **8**, 737347 (2021).

25. W. Yan, *et al.*, *Phys. Rev. B* **97**, 205422 (2018).

26. Y. Kang *et al., Phys. Rev. B* **78**,115404 (2008).

27. M. L. Brongersma *et al., Nat. Nanotechnol.* **10**, 25−34 (2015).

28. M. B. Lundeberg *et al., Nat. Mater.* **16**, 204−207 (2017).

29. F. Wang *et al., Nat. Commun.* **14**, 2224 (2023).

30. X. Cai *et al., Nat. Nanotechnol.* **9**, 814−819 (2014).

31. S. M. Koepfli *et al., Science* **380**, 1169−1174 (2023).

32. L. Viti *et al., Nano Lett.* **20**, 3169−3177 (2020).

33. N. Tarjányi *et al., Opt. Mater.* **37**, 798−803 (2014).

34. B. H. Kim *et al., Opt. Lett.* **6**, 1657−1659 (2001).

35. T. Low *et al., Nat. Mater.* **16**, 182−194 (2017).

36. Z. Zheng *et al., Sci. Adv.* **5**, eaav8690 (2019).

37. F. Wang *et al., Nat. Commun.* **12**, 5628 (2021).

38. J. Taboada-Gutiérrez *et al., Nat. Mater.* **19**, 964−968 (2020).

**Acknowledgments:**

**Funding:**

The National Key Basic Research Program of China grants 2019YFA0210200, 2019YFA0210203, 2022YFA1203500, and 2022YFA1206600

The National Natural Science Foundation of China grant 91963205

The Changjiang Young Scholar Program

**Author contributions:**

Conceptualization: SZD, HJC, and NSX

Methodology: HJC, SZD, XMW, NSX, SJL, HJZ, and ZLC

Investigation: HJC, SZD, NSX, XMW, SJL, HJZ, and ZLC

Visualization: XMW, SJL, HJC, SZD, NSX, and HJZ

Funding acquisition: SZD, NSX, and HJC

Project administration: HJC, SZD, and NSX

Supervision: NSX, SZD, and HJC

Writing – original draft: HJC, SZD, NSX, and XMW

Writing – review & editing: all authors

**Competing interests:** Authors declare that they have no competing interests.

**Data and materials availability:** All data are available in the main text or the supplementary materials.

**Supplementary Materials**

Materials and Methods MM1 to MM4

Supplementary Text ST1 to ST6

Figs. S1 to S17

Tables S1

References (1–15)

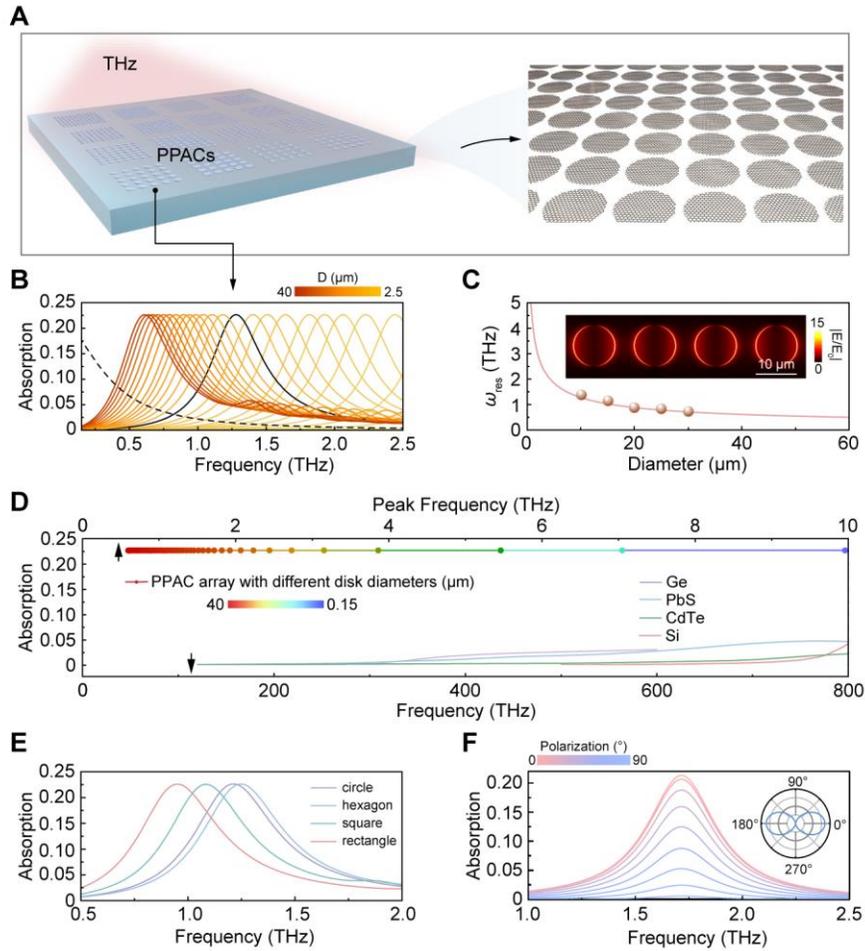

**Fig. 1. Plasmon polariton resonances of the PPACs in the THz spectral range.** (**A**) Schematic showing the PPACs. (**B**) Theoretical THz absorption spectra of PPACs of disk shape. The graphene disks have diameters of 2.5 μm to 40 μm. The absorption spectrum of an unpatterned monolayer graphene flake is also depicted for comparison (dashed line). (**C**) Dependence of the resonance frequency of the graphene disks on their diameters. Solid line: theoretical results. Symbols: experimental results extracted from the experimental extinction spectra. Inset: simulated electromagnetic wave localizations of four graphene disks at the PPAC resonance frequency of 1.24 THz. (**D**) Comparison of the electromagnetic absorption characteristics between typical semiconductors and disk-shaped PPAC arrays. The thicknesses of the semiconductors are set as 0.7 nm, the same as that of a monolayer graphene. For the semiconductors their absorption spectra covering the corresponding electron bandgaps are given. (**E**) Theoretical absorption spectra of PPACs with different shapes. The diameter of the disk, side lengths of the hexagon and square, and length of the ribbon are all 10 μm. (**F**) Theoretical THz absorption spectra of a nanoribbon PPAC at different excitation polarization angles. The polarization angle is defined as the angle between the polarization direction of the incident THz wave and the ribbon length axis. Inset: polar plot of the absorption maximum against the polarization angle, which displays a cosine function behavior.

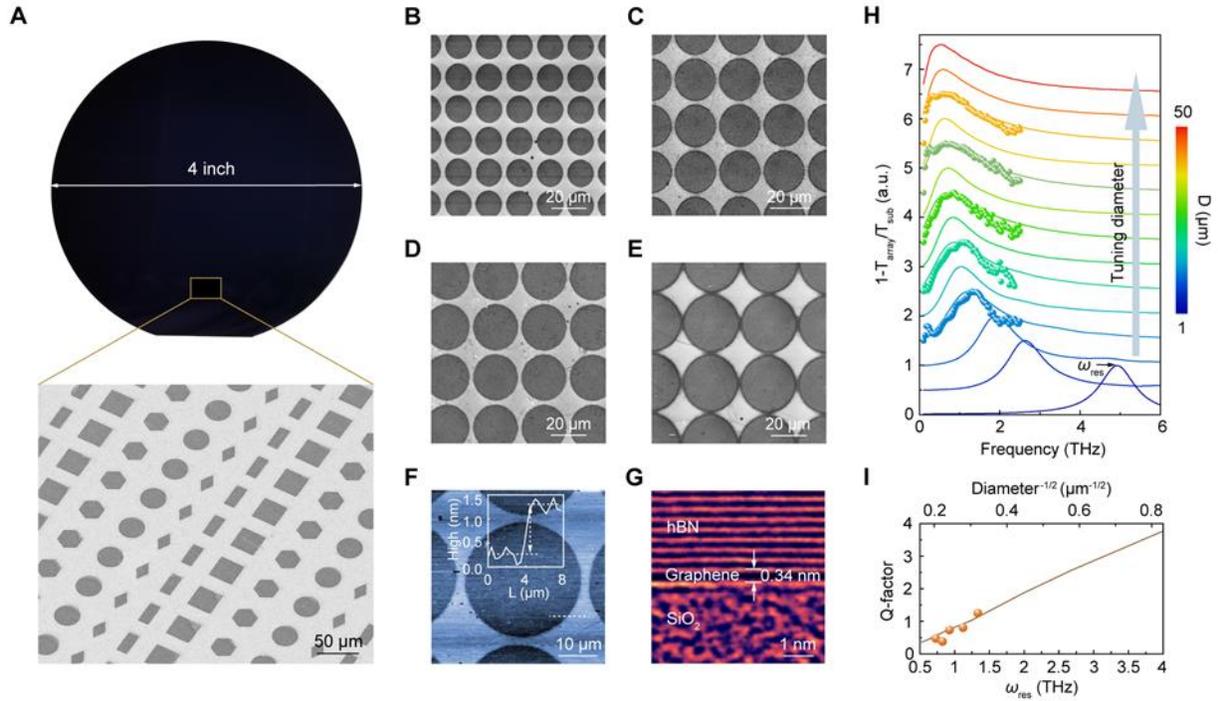

**Fig. 2. Fabrication and characterization of the THz PPACs.** (**A**) Photograph (upper panel) and enlarged scanning electron microscope image (lower panel) of a 4-inch PPACs wafer. (**B** to **E**) Scanning electron microscope images of four typical circular-shaped PPACs of different diameters. (**F**) Atomic force microscope image of a representative PPAC structure. Inset is the topology profile taken along the dashed white line across the disk edge, showing a thickness of 0.97 nm. (**G**) Cross-sectional transmission electron microscope image of a representative PPAC structure, showing the thickness of a single carbon atom layer. The difference between the thicknesses measured from the atomic force microscope and transmission electron

microscope is due to the wrinkle formed on graphene supported onto the substrate. (**H**) Theoretical and experimental extinction spectra of the arrays composed of circular PPACs of different diameters. Solid lines: theoretical results. Symbols: experimental data. (**I**) Dependence of quality factor (Q-factor) on resonance frequency $\omega_{res}$ extracted from the simulated (solid line) and experimental (symbols) absorption spectra shown in (**H**). The Q-factor as a function of $D^{-1/2}$ is also given (upper axis).

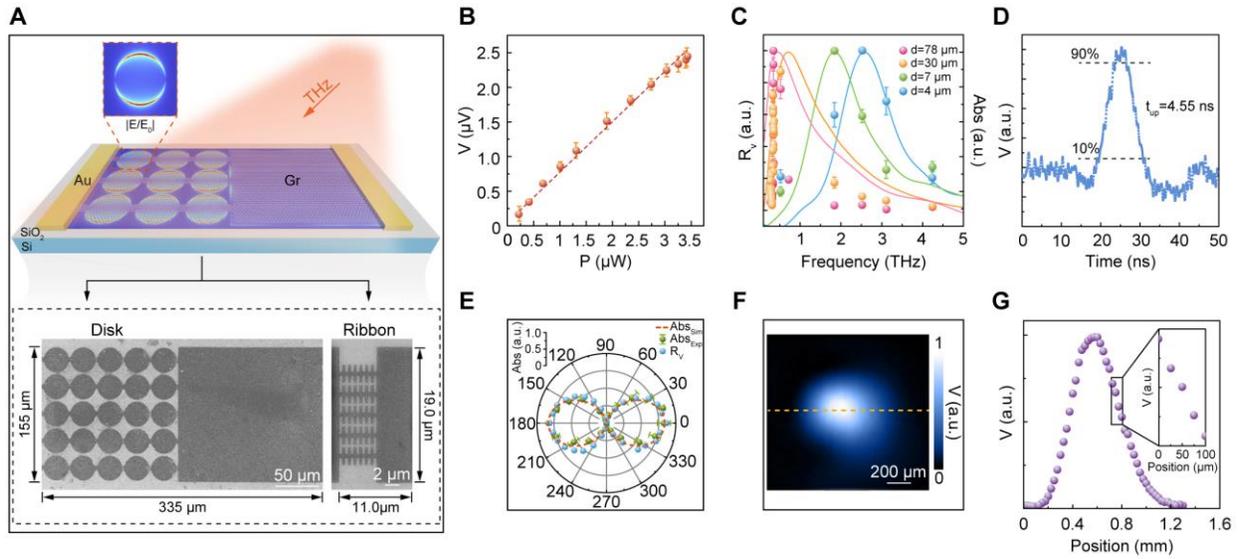

**Fig. 3. PPAC detectors and their THz detection performances.** (**A**) Schematic illustrating the configuration of the THz detector. The active layer consists of a monolayer graphene, with the left region comprising PPACs interconnected by graphene strips. The right region remains unpatterned. THz near-field distribution at the resonance frequency is overlaid on the PPACs and unpatterned graphene region, showing the strong electromagnetic field localizations at the PPACs and their edges. Lower panel: scanning electron microscope image of a detector composed of disk-shaped (left) and ribbon-shaped (right) PPACs. The disk diameter is 30 μm. The spacing between different disks is 2 μm. The graphene strip width is 5 μm. Each ribbon has a length and width of 2.5 μm and 0.5 μm, respectively. The spacing between the edges of adjacent ribbons is 1 μm. The graphene strip width is 1 μm. (**B**) Photo-

induced voltage as a function of the THz illumination power. Symbols: experimental data. Dashed line: linear fit to the symbols. (**C**) Dependence of the photo-induced voltage responsivity $R_v$ on the THz frequency (symbols, left vertical axis). The colored lines (right vertical axis) show the calculated absorption spectra of the PPACs, which are normalized for comparison. The colors delegate detectors composed of PPACs with different diameters. The orange ones correspond to device composed of disk-shaped PPACs shown in (**A**). (**D**) Photo-induced voltage of the disk-shaped PPAC detector shown in (**A**) in response to a THz pulse (0.1−3.0 THz) illumination. (**E**) Polar plots of the photo-induced voltage responsivity (blue symbols), simulated absorption intensity (orange dashed line), and experimental absorption intensity (green symbols) as a function of the incident THz polarization angle. The polarization angle is defined as the angle between the polarization direction of the incident THz wave and the ribbon length axis. The illumination frequency is 0.34 THz. (**F**) Photo-induced voltage mapping of the THz laser spot. The image is obtained by raster-scanning the detector across the focus of the laser beam at 2.52 THz. The polarization of the THz laser is parallel to the length axis of the graphene nanoribbon. (**G**) Line profile of photo-induced voltage across the focused laser spot along the dashed yellow line shown in (**F**). The line profile indicates that intensity differences between adjacent positions, with a separation of 25 μm, can be distinguished. All detectors are fabricated onto silicon substrate covered with a 300 nm-thick SiO$_2$ layer.

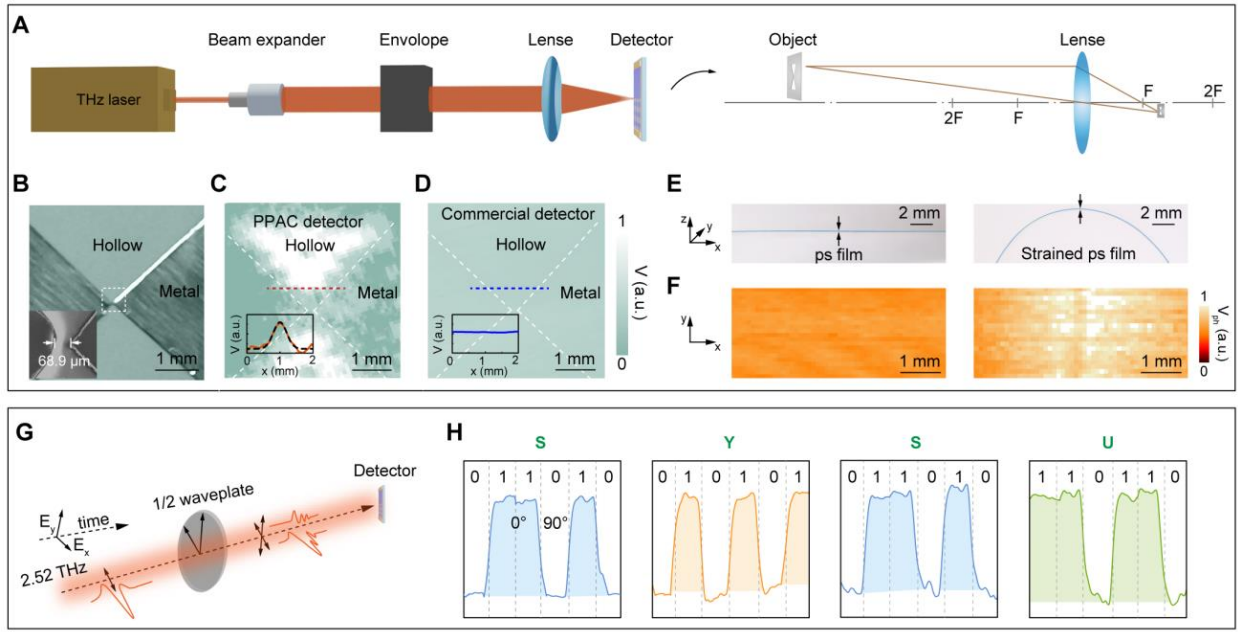

**Fig. 4. Proof-of-concept applications of the PPAC detectors.** (**A**) Scheme of THz stealth imaging of an object using the PPAC detector. The image plane is positioned very close to the focal plane of the objective lens due to the nearly parallel illumination of the object with a THz plane wave. (**B**) Optical image of a metallic plate with a hollow bowtie structure at its center, which is placed inside an opaque envelope. Inset: scanning electron microscope image corresponding to the region enclosed by the dashed square, showing a gap size of 68.9 μm. (**C**) THz image of the hollow bowtie structure acquired using the PPAC detector. Inset: photo-induced voltage line profile along the red dashed line, where the dashed black line indicate the Gaussian fit of the line profile. (**D**) THz image of the same metallic plate corresponding to (**C**) using a commercial detector, where only featureless intensity distribution can be obtained. Inset: photo-induced voltage line profile along the blue dashed line. (**E**) Optical images of the polyester films in free state (left) and subjected to strain (right). (**F**) Photo-induced voltages of the polarization-sensitive PPAC detector in response to a 2.52-THz wave transmitting through a polyester film in free state (left) and strain (right). The polarization the THz wave is set at 45° with respect to the length axis of the graphene nanoribbon in the PPAC detector. The flexible polyester film is placed inside an envelope. (**G**) Scheme of the experimental setup for polarization multiplexed THz wireless communication. The polarization of the 2.52-THz laser is switched between the two orthogonal directions by rotating the half-wave plate. Subsequently, the laser is guided and focused onto the polarization-sensitive PPAC detector. A specific string, such as "SYSU", was first converted into ASCII codes and then encoded onto the dual polarization states of a 2.52-THz carrier signal. The polarization angle is determined by the orientation between the laser's polarization direction and the length axis of the graphene nanoribbon in the PPAC detector. (**H**) Photo-induced voltages of the polarization-sensitive PPAC detector in response to the THz wave carrying the ASCII codes of the string "SYSU", showing exact correlation with the original encoded information.